# The valleytronic topological filters in silicene-like inner-edge systems

Xie Hang (谢航)[1,2], Lü Xiao-Long (吕小龙)[3*] and Yang Jia-En (杨加恩)[1,4†]

[1]*College of Physics, Chongqing University, Chongqing, China, 401331*

[2]*Chongqing Key Laboratory for Strongly-Coupled Physics, Chongqing University, Chongqing, China, 401331*

[3]*College of Science, Guangxi University of Science and Technology, Liuzhou, China，545006*

[4]*School of Electronics and IoT, Chongqing College of Electronic Engineering, Chongqing, China，401331*

[*]Corresponding author, Email:   physicslxl@163.com
[†]Corresponding author, Email:   yangjiaen309@163.com

Inner edge state with spin and valley degrees of freedom is a promising candidate to design a dissipationless device due to the topological protection. The central challenge for the application of inner edge state is to generate and modulate the polarized currents. In this work, we discover a new mechanism to generate fully valley- and spin-valley-polarized current caused by the Bloch wavevector mismatch (BWM). Based on this mechanism, we design some serial-typed inner-edge filters. With once of the BWM, the coincident states could be divided into transmitted and reflected modes, which can serve as a valley or spin-valley filter. In particular, while with twice of the BWM, the incident current is absolutely reflected to support an off state with a specified valley and spin, which is different from the gap effect. These findings give rise to a new platform for designing valleytronics and spin-valleytronics.

## 1. Introduction

As an essential two-dimensional material, graphene is vital in designing novel devices due to the controllable linear Dirac dispersion [1, 2]. However, graphene's planar structure and weak spin-orbit interaction limit its application. Different from graphene, silicene-like materials, such as silicene, germanene, stanene, and other silicene-derived materials [3-5] with a buckled honeycombed structure and a stronger spin-orbit interaction, are important platforms to investigate the fundamental quantum transport physics and for a wide range of applications in designing low dissipation devices. It is found that silicene-like materials have two inequivalent valleys (called K and K') in the first Brillouin zone. Similar to the spin degree of freedom in spintronic devices, these inequivalent valleys or valley degrees of freedom can be used as information carriers to give rise to valleytronics [6-8]. Many unique transport phenomena can be realized by controlling these two degrees of freedom, such as the valley filters, fully spin-polarized currents and spin-valley filters [9-13].

Silicene-like materials can exhibit abundant topological phases including quantum anomalous Hall (QAH), quantum spin Hall effect (QSH), quantum valley Hall (OVH), quantum spin-valley Hall (QSVH), and spin QAH under some external fields [14-16], such as electric field, side potential, off-resonant circularly polarized light and antiferromagnetic field applied on the whole system. According to the bulk-edge correspondence, these topological phases correspond to their outer edge states. In addition, researchers have also found a large amount of inner-edge states mainly distributed at the interface separating different topological phases [17-20]. By applying a spatially varying external field in zigzag graphene-like materials, the valley-polarized and spin-valley-polarized currents can be obtained due to the valley-momentum and spin-valley-momentum locked conducting channels, which can be regarded as the valley and spin-valley filters [9-13]. In addition, the double-output and single-output spin-valley filters can be generated by manipulating the spin-valley polarized current in a three-terminal device [19]. The conductance of the inner-edge states is also robust to various types of long- and short-range perturbations [15, 21]. It should be noted that kink states, also known as valley inner-edge states, occur in graphene systems [22-25] or photonic crystals [26, 27].

Though the inner-edge states have been extensively investigated, some critical points are still worth exploring in depth. For example, most inner-edge states focus on the combination of different QVH states[10, 25], QSH and QVH states[15], or QAH and QVH states[28]. However, other combinations of topological phases are still needed to be investigated. Moreover, most researchers only work with pure one-junction structures; the serial multi-segments structures can generate much ampler physical phenomena with new valley and spin filtering effects.

In this work, we report the transport properties of the inner-edge states in QAH, QVH and SQVH phases based on the hybrid silicene-like nanoribbons, which are valley or spin-valley polarized. A novel filtering mechanism called the Block Wavevector Mismatch, which has not been reported before, has been found. The two-segments and three-segments serial inner-edge filters have been designed using this mechanism. These filters are capable of producing valley-polarised or valley-spin-polarised inner-edge currents. Our analysis shows that these new inner-edge filters could potentially serve as promising components for future valleytronic devices.

## 2. Models and Theories

### 2.1 The model of hybrid silicene-like nanoribbons

In this work, we construct some hybrid silicene-like nanoribbons for the valley or spin filter. This hybrid nanoribbon consists of two adjacent nanoribbons with different topological phases (Fig. 1(a)). The inner-edge state may exist near the interface of these two adjacent nanoribbons. We also serially combine two or three segments of these hybrid nanoribbons to form some filters with different valley or valley-spin polarization functions (Fig. 1(b) and 1(c)).

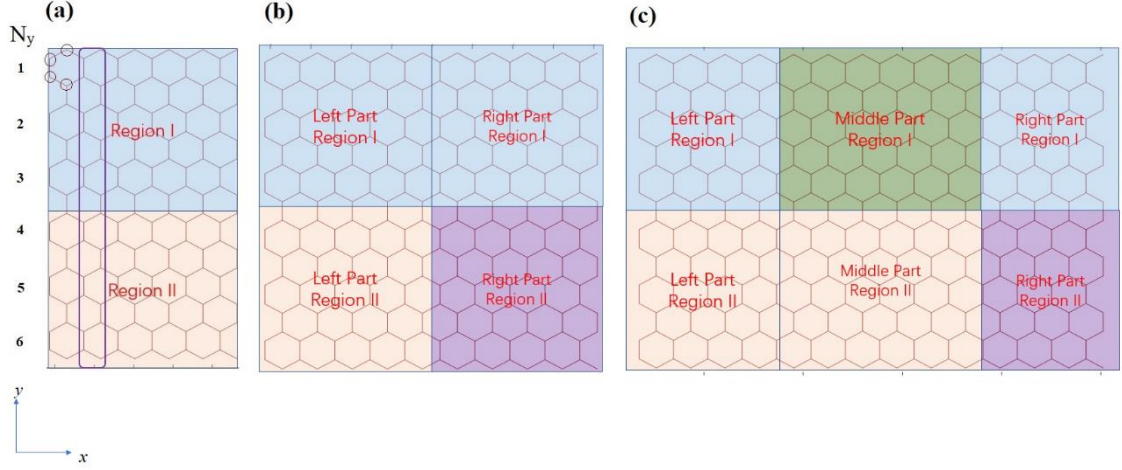

**Fig. 1.** (a) The hybrid silicene-like nanoribbon. The upper and lower parts are labeled region I and region II. The rectangle denotes the periodic unit cell with $N_y$ four-atom sub-units; (b) The two-parts valley(-spin) filter based on the hybrid silicene-like nanoribbons; (c) The three-parts valley-spin filter based on the hybrid silicene-like nanoribbons.

These hybrid silicene-like nanoribbons are described as the tight-binding model with the nearest neighbor approximation. The Hamiltonian of the system are shown below[20].

$$H = t\sum_{<i,j>}\sum_{\alpha} c_{i\alpha}^{\dagger} c_{j\alpha} + it_{SO}\sum_{<<i,j>>}\sum_{\alpha} v_{ij} s_{\alpha} c_{i\alpha}^{\dagger} c_{j\alpha} + i\sum_{<<i,j>>}\sum_{\alpha} t_{\Omega,i} v_{ij} c_{i\alpha}^{\dagger} c_{j\alpha} + \sum_{i}\sum_{\alpha} V_i u_i c_{i\alpha}^{\dagger} c_{i\alpha} \\ + \sum_{i}\sum_{\alpha} M_i u_i s_{\alpha} c_{i\alpha}^{\dagger} c_{i\alpha} \qquad (1)$$

The first term is the hopping term between the site $i$ and $j$ with the spin $\alpha$ ($\alpha = \uparrow; \downarrow$), and $t$ is the hopping energy, which is different for silicene (1.6 eV), germanene (1.3 eV) and stanene (1.3 eV) [4]. <,> means the nearest neighbor sites the summation. The second term is the intrinsic spin-orbital interaction with the coupling energy $t_{SO}$, which is usually much smaller than $t$. <<,>> means the summation is implemented between the next nearest neighbor sites. $v_{ij} = \pm 1$ when the hopping path from site $i$ to $j$ is clockwise (anti-clockwise) with respect to the positive $z$ axis. $s_{\alpha}$ is the spin factor, $s_{\alpha} = \pm 1$ for $\alpha = \uparrow$ or $\downarrow$. The third term is for the Haldane light field with the site-dependent intensity $t_{\Omega,i}$. This intensity may be positive or negative, depending on the light radiated on the nanoribbon is left or right circularly-polarized. The fourth term is for the staggered electric potential with the site-dependent electric potential $V_i$, and $u_i = \pm 1$ for site $i$ in A or B sublattice. This staggered electric potential can be fulfilled by the vertical electric field exerted on the nanoribbon plane due to the buckled structure of silicene-like materials. The last term is the antiferromagnetic (AF) exchange interaction with the site-dependent intensity $M_i$. This is implemented by absorbing the magnetic atoms, or sandwiching the non-magnetic perovskite structure by two perovskite layers [29, 30].

Here we give more descriptions for this AF term. With the proximity coupling of ferromagnetic materials, the ferromagnetic (FM) exchange field also exists in SiNR. This can be employed by depositing silicene to a ferromagnetic insulating substrate. FM exchange field provides the Zeeman effect to different spin components, and can generate the spin-polarized current [31-33]. But due to the buckled structure of SiNR, the A/B sublattices feel different FM fields. The anti-symmetric part of these different FM fields ($(M_{FM,A} - M_{FM,B})/2$) can give rise to the AF exchange field, which contributes to the SQVH topological state in this study. The detailed generation and effects of FM and AF exchange fields will be investigated in the further study.

In our hybrid nanoribbons, we modulate the signs of the site-dependent intensities $t_{\Omega,i}$, $V_i$ and $M_i$ when site $i$ is in region I or region II (see Fig. 1). This can result in some inner-edge states. Each region in the part of the filter has uniform external fields. From the Dirac equation of 2D materials and the corresponding topological insulator theory, we know that different combinations of external field parameters can lead to different topological phases and different types of inner-edge states. This can be shown in the following section.

## 2.2 Dirac equation and the Jackiw-Rebbi interface mode

In the Hamiltonian above, with the new parameters $t_{SO} = \lambda_{SO}/3\sqrt{3}$ and $t_{\Omega} = \lambda_{\Omega}/3\sqrt{3}$, and using the Fourier transformation to transform the real space quantity into the reciprocal space (here we consider the uniform external fields), we may obtain

the Dirac equation in the k-space near the two Dirac points (K and K' points)

$$H_D = \hbar v_F (k_x \sigma_x + \eta k_y \sigma_y) + (\eta s \lambda_{SO} + \eta \lambda_\Omega + sM + V)\sigma_z, \qquad (2)$$

where $(k_x, k_y)$ is the wave vector derivation from the Dirac point K or K', $\eta = \pm 1$ is for the K or K' valley. $s = \pm 1$ stands for the two spin (up and down) cases, $\sigma_x$, $\sigma_y$ and $\sigma_z$ are the Pauli matrices in three directions. $M$ and $V$ are energy of the AF exchange field and staggered electric potential. We also call the last term above as the mass term[14]

$$m_{\eta s} = \eta s \lambda_{SO} + \eta \lambda_\Omega + sM + V. \qquad (3)$$

It is known that for this Dirac periodic system, the non-zero mass term leads to the nonzero Berry curvature, and the integral of this Berry curvature in the first Brillouin zone is the Chern number of this system, which read

$$C_{\eta s} = \frac{1}{2} \text{sgn}(m_{\eta s}). \qquad (4)$$

Since the Chern number is related to the topological properties (The detailed relation can be found in the tables of the previous work[14, 20]), we see that the sign of the mass term determines which topological phase the system belongs to. From Eq. (3) we know that with proper combinations of the external fields, we may modulate the signs of the mass term, or the topological phase of the system. Thus there exist some phase diagrams [20], which can quickly tell us the phase region for these parameter combinations (such as $\lambda_\Omega \sim M$ or $\lambda_\Omega \sim V$). Our work is based on these phase diagrams.

Now we investigate the inner-edge state of the Dirac system. Considering the system with two adjacent uniform external potentials along a straight line, the wavefunction can be written as the guide-wave form: $\Psi(x,y) = e^{ik_x x} \begin{pmatrix} \phi_1(y) \\ \phi_B(y) \end{pmatrix}$. Substitute this into the Dirac equation (2), and with the transform $(k_x, k_y) \Rightarrow (\partial_x, \partial_y)$, the following differential equations are obtained

$$\begin{cases} m_{\eta s}(y)\phi_1(y) + \hbar v_F(k_x - \eta\partial_y)\phi_2(y) = \varepsilon\phi_1(y) \\ \hbar v_F(k_x + \eta\partial_y)\phi_1(y) - m_{\eta s}(y)\phi_2(y) = \varepsilon\phi_2(y) \end{cases} \quad (5)$$

Then we get the equation of $\phi_1(y)$:

$$(\hbar v_F)^2 \partial_y^2 \phi_1(y) - [m_{\eta s}^2(y) - \varepsilon^2 - (\hbar v_F k_x)^2]\phi_1(y) = 0 \quad (6)$$

Near the valley-Fermi-energy region ($k_x \approx 0$; $\varepsilon \approx 0$) [20], $\phi_1(y)$ has the solution with the form $\phi_1(y) = C_1 e^{\int^y m_{\eta s}(y')dy'} + C_2 e^{-\int^y m_{\eta s}(y')dy'}$. With the step-wised $m_{\eta s}(y)$ v.s. $y$ curve, it is proved that only when $m_{\eta s}$ has different signs in the two uniform potentials (Region I and II), there exists an inner-edge state (Jackiw-Rebbi mode) which decays exponentially apart from the interface line ($y=0$)[14]:

$$\phi_1(y) = \begin{cases} C_1 e^{\frac{1}{\hbar v_F} m_{\eta s}^{II} y} & (y < 0) \\ C_2 e^{-\frac{1}{\hbar v_F} m_{\eta s}^{I} y} & (y \geq 0) \end{cases} \quad (7)$$

And the slope of the inner-edge band is determined by the Chern number difference [14, 18]

$$\Delta C_{\eta,s} = C_{\eta,s}^I - C_{\eta,s}^{II}. \quad (8)$$

So we can use this $\Delta C_{\eta,s}$ to determine the property of the inner-edge band with specified valley and spin. This Chern number difference is also called the inner-edge Chern number. The detailed discussions of these inner-edges are in Section 3.

**2.3 The transport theory: non-equilibrium Green's function**

To further investigate the inter-valley and intra-valley scatterings of inner-edge states constructed to a two-terminal device, the transmission coefficient based on non-equilibrium Green's function (NEGF) theory is introduced as

$$T^s = \text{Tr}\left[\boldsymbol{\Gamma}_R^s \mathbf{G}^{r,s} \boldsymbol{\Gamma}_L^s \mathbf{G}^{a,s}\right]. \quad (9)$$

This formula describes the spin-dependent (denoted by the superscript $s$) transmission from the left lead (denoted as $L$) to right lead (denoted as $R$). And the linewidth function is defined as $\boldsymbol{\Gamma}_{L/R}^s = i\left(\boldsymbol{\Sigma}_{L/R}^{r,s} - \boldsymbol{\Sigma}_{L/R}^{a,s}\right)$, where the advanced and retarded self-energy functions can be obtained from the surface green's function $\mathbf{g}_{L/R}^{r/a,s}$ of the leads and the two coupling matrices $\mathbf{H}^s_{D,L/R}$ and $\mathbf{H}^s_{L/R,D}$: $\mathbf{\Sigma}^{r/a,s}_{L/R} = \mathbf{H}^s_{D,L/R}\mathbf{g}^{r/a,s}_{L/R}\mathbf{H}^s_{L/R,D}$. Meanwhile, the advanced and retarded Green's functions in Eq. (9) are also dependent on their corresponding self-energy functions, which is written as $\mathbf{G}^{r/a,s} = \left[(E\pm i\delta)\mathbf{I} - \mathbf{H}^s_D - \mathbf{\Sigma}^{r,s}_L - \mathbf{\Sigma}^{r,s}_R\right]^{-1}$. It is obvious that the surface Green's function $\mathbf{g}^{r/a,s}_{L/R}$ of the leads is the central issue for calculating the transmission, which can be calculated by the Lopez–Sancho's iterative method or transfer matrix.

To intuitively exhibit the transmitted and reflected channels, we need to calculate the local bond currents which play an important role in discovering novel transport phenomena and capturing the transport path of inner edge state. We here introduce the energy-and-spin-resolved local bond current from the site $i$ to $j$, expressed as[19]

$$J^s_{ij}(E) = H^s_{ji}G^{<,s}_{ij}(E) - H^s_{ij}G^{<,s}_{ji}(E), \tag{10}$$

where $H^s_{ji}$ is the Hamiltonian matrix element and $\mathbf{G}^{<,s} = -i\mathbf{G}^{r,s}\mathbf{\Gamma}^s_L\mathbf{G}^{a,s}$ is the lesser Green's function based on the suppose that the incident current is from the left lead.

## 3. Results and Discussions

In this section, we first introduce some types of topological inner-edges states in the hybrid silicene nanoribbons. The Chern number analysis is employed to relate the topological property with the valley, spin, and momentum direction of these inner-edge states. Then we combine two or three segments of these hybrid nanoribbons to construct the valley or spin-valley filters. The local current distributions plus the energy band analysis are employed to reveal the filtering mechanism of the inner-edges or outer-edge device.

In the following calculations, we use these parameters: the intrinsic SOC intensity is set as $t_{SO} = 0.01t$, the circularly-polarized light intensity is $t_\Omega = \pm 0.3t$ for the QAH1 and QAH2 phases; the staggered potential energy is set as $V = \pm 0.3t$ for the QVH1 and QVH2 phases; and the AF exchange intensity is set as $M = \pm 0.3t$ for the SQVH1 and SQVH2 phases. The detailed phase diagram information can be referred in our previous work[20].

### 3.1 Inner and outer edge states in hybrid nanoribbons

Here we investigate three types of topological hybrid nanoribbons with their band structures and the eigenmode distributions of the inner-edge and outer-edges. We also use the Jackiw-Rebbi interface theory to analyze the topological properties of the interface transport.

**(1) QAH1|QAH2 case**

In this case, two TI regions across the inner-edge are the QAH states with different Chern numbers: in one region $C_{total}=2$ (QAH1) and in the other region $C_{total}=-2$ (QAH2). The detailed spin and valley-resolved Chern numbers are also listed in Table 1 below. These two topological phases can be implemented by the radiation of the right (left)- circularly polarized light with the off-resonance frequency [14, 34].

**Table 1.** The Chern numbers of the hybrid nanoribbon system with QAH1|QAH2 phases

| Topological Region | $2C_{K,\uparrow}$ | $2C_{K',\uparrow}$ | $2C_{K,\downarrow}$ | $2C_{K',\downarrow}$ |
|---|---|---|---|---|
| I: QAH1 | 1 | 1 | 1 | 1 |
| II: QAH2 | -1 | -1 | -1 | -1 |
| Interface: $2(C^I-C^{II})$ | 2 | 2 | 2 | 2 |

From the Jackiw-Rebbi theory in Section 2, we know the topological property are determined by the difference of two Chern numbers, which are listed in Table 1. The band structures of this hybrid nanoribbon with two spins are shown in Fig. 2 (a) and 2(b). We see that they are much like spin degenerate bands. Two single bands connect the valence and conductance bulk bands with the positive slope in the K and K' valley regions. From the electron distributions of these two edge bands (Fig. 2(c) and (d)), we find they correspond to the inner-edge of the hybrid nanoribbon. In the Jackiw-Rebbi theory, the momentum direction of the inner-edges with corresponding spin and valley agrees well with the sign of interface Chern number $\Delta C_{\eta,s} = C^I_{\eta,s} - C^{II}_{\eta,s}$. That means if $\Delta C_{\eta,s}$ is positive (negative), the momentum direction of inner-edge with the valley $\eta$ and spin $s$ is also positive (negative). In Fig. 2(a) and 2(d), we see that the slopes of these four inner-edge bands (with K and K' valley and up and down spins) are positive, which is agreeable with the four Chern numbers ($\Delta C_{K,\uparrow}, \Delta C_{K',\uparrow}, \Delta C_{K,\downarrow}, \Delta C_{K',\downarrow}$) of the inner-edge state (see Table 1).

In Fig.2(a) and(b), there also exist two groups of edge bands with negative slope, which connect the valence K' valley and conductance K valley bulk bands. Each group contains two nearly degenerate bands and these bands all correspond to the outer-edge state, as shown in Fig. 2(e) and 2(f). One is near the upper edge and another is near the lower edge. In the band figures we see that the inner-edge states (in two valleys) are with positive group velocities, which are balanced by the out-edge states (near two edges) with negative group velocities.

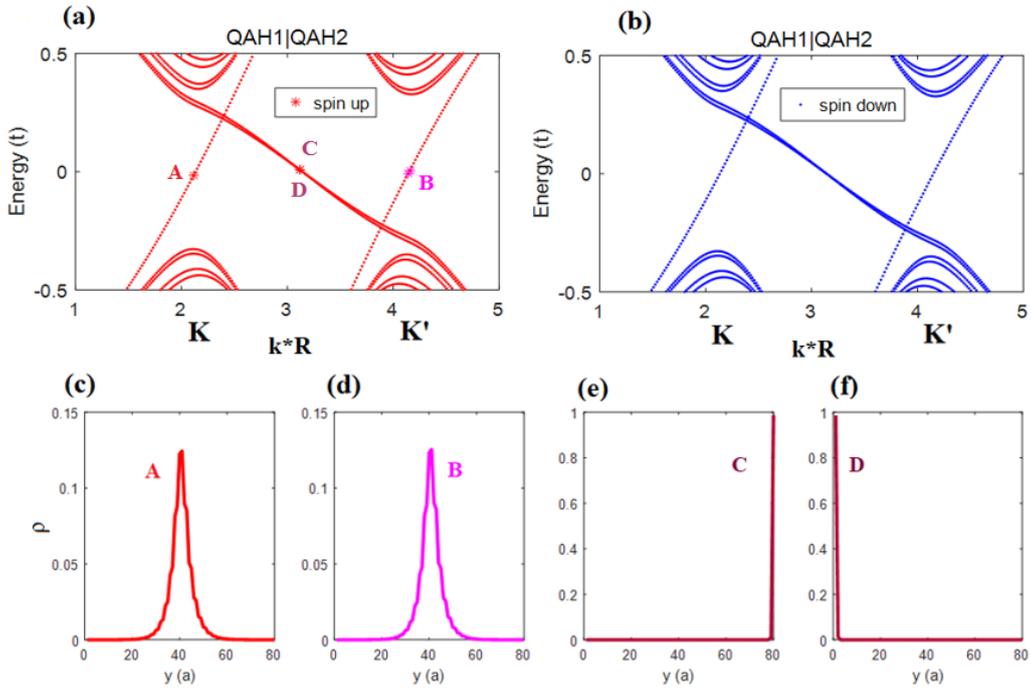

**Fig. 2.** The band structures and electron distributions of inner/outer edges in the hybrid nanoribbon (QAH1|QAH2). (a) and (b) are for the energy bands for spin-up and spin-down case respectively; (c) and (d) are the inner edges in the K and K' valley in spin-up case; (e) and (f) are the outer edges near the energy E=0 for the two nearly degenerate bands (with up spin). Their corresponding states are dotted and labeled as 'A', 'B', 'C' and 'D' in (a).

**(2) QAH|QVH case**

In this case, the hybrid nanoribbon contains two topological phases: QAH and QVH phases. Since these bands are nearly spin degenerate (for silicene nanoribbon, the difference is neglectable for two spin bands), in Table 2 we only list the Chern numbers with different valley indices. In region I the topological phase is QAH1; In region II we investigate two types of QVH: QVH1 and QVH2 (denoted as 'A' and 'B' ). From Table 2 the Chern numbers of the inner-edge state for type A are $2\Delta C_{K,\uparrow(\downarrow)}=0$ and $2\Delta C_{K',\uparrow(\downarrow)}=-2$. That means only in K' valley there is an inner-edge band with a negative slope (Fig3. (a)). Similarly in the type B (QAH1|QVH2) case $2\Delta C_{K,\uparrow(\downarrow)}=2$ and $2\Delta C_{K',\uparrow(\downarrow)}=0$, there is an inner-edge band in K valley with a positive slope, as shown in Fig3. (b).

**Table 2.** The Chern numbers of the hybrid nanoribbon system with QAH|QVH phases

| Topological Region | $C_{K,\uparrow(\downarrow)}$ | $C_{K',\uparrow(\downarrow)}$ |
| --- | --- | --- |
| I: QAH1 | 1 | 1 |
| II: QVH1(A)  or  QVH2(B) | 1 (A)  or  -1 (B) | -1(A)  or  1(B) |
| Interface: $2(C^I-C^{II})$ | 0 (A)  or  2 (B) | -2(A)  or  0(B) |

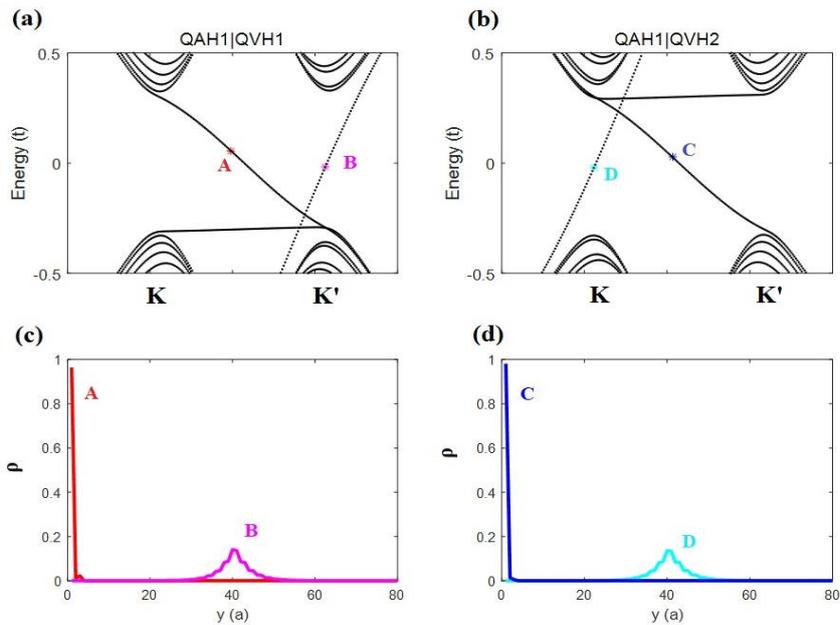

**Fig. 3.** The band structures ((a)) and electron distributions of inner/outer edges ((c)) in the hybrid nanoribbon of QAH1|QVH1 case; The band structures ((b)) and electron distributions of inner/outer edges ((d)) in the hybrid nanoribbon of QAH1|QVH2 case. The corresponding states in (c) and (d) are dotted and labeled as 'A', 'B', 'C' and 'D' in (a) and (b). In (a) and (b) the two bands are spin-degenerate, so we use black curves to stand for two spins.

In Figs. 3(c) and 3(d), the inner-edge electron distributions are displayed. The corresponding eigenstate points are dotted and labelled as 'B' and 'D' in the band figures 3(a) and 3(b). Besides, in the band figures there still exist the outer-edge bands connecting K' valance and K conductance bulk bands (dotted and labelled as 'A' and 'C'). The corresponding outer-edge electron distributions are presented in the lower panels. We also observe that all the outer-edges are located on the upper edges.

### (3) QAH|SQVH case

In this case, we use QAH1 and SQVH1 as the two topological phases in region I and II respectively. The corresponding Chern numbers of the two regions and the interface are listed in Table 3 and the band structures are shown in Fig. 4 (a) and (b). Similar to the QAH1|QVH1 case, there is only one inner-edge band in one valley. But the inner-edge band of spin up (down) polarization locates in K' (K) valley. We see that here the valley and spin of the inner-edge are locked with each other, due to the SQVH topological phase. These inner-edge bands are also agreeable with their Chern numbers: $2\Delta C_{K,\uparrow} = 0$; $2\Delta C_{K',\uparrow} = 2$; $2\Delta C_{K,\downarrow} = 2$; $2\Delta C_{K',\downarrow} = 0$.

**Table 3.** The Chern numbers of the hybrid nanoribbon system with QAH1|SQVH1 phases

| Topological Region | $2C_{K,\uparrow}$ | $2C_{K',\uparrow}$ | $2C_{K,\downarrow}$ | $2C_{K',\downarrow}$ |
|---|---|---|---|---|
| I:QAH1 | 1 | 1 | 1 | 1 |
| II:SQVH1 | 1 | -1 | -1 | 1 |
| Interface: $2(C^{I}-C^{II})$ | 0 | 2 | 2 | 0 |

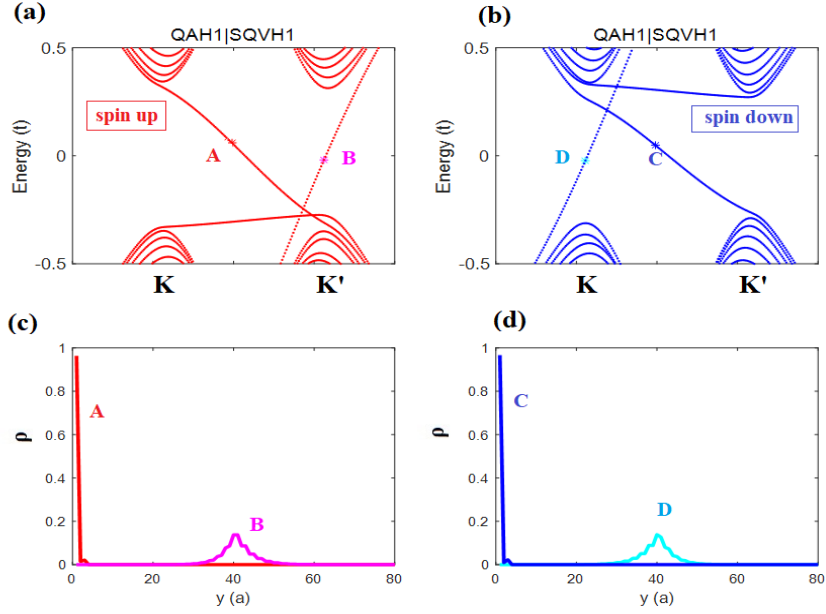

**Fig. 4.** The band structures and electron distributions of inner/outer edges in the hybrid nanoribbon with the QAH1|SQAH1 topological phases. (a) and (c) are for spin up case and (b) and (d) are for spin down case. Labels 'A' and 'C' stand for the outer edges and labels 'B' and 'D' stand for the inner edges.

**3.2 Valley-polarized transport in hybrid nanoribbon junctions**

Now we combine the two different hybrid nanoribbons to obtain the single valley filter as shown in Fig. 5 below. The topological structure of this filter is in Fig. 5(a): the left part is the combination of QAH1|QAH2, with the inner-edge in K and K' valleys. The corresponding band structure of QAH1|QAH2 is shown in Fig. 2. And the right part of the filter is the combination of QAH1|QVH2 with a single valley-polarized (K) inner-edge (the corresponding band structure is in Fig. 3).

Since there are two valley channels in the inner current of the left part but only a single valley channel in the right part, we expect that the K' valley component will be filtered. This is validated by our calculated local current distribution. As shown in Fig. 5(b), the inner-edge current flows rightward to the interface of two hybrid nanoribbons and the K'-valley-polarized component are blocked and then it goes upward along the vertical interface to the upper edge. This block of K'-valley current is due to the BWM effect, which is much different from the traditional band gap effect in the filtering mechanism. Since there are outer-edge bands (with negative slope) in the QAH1|QAH2

hybrid system (Fig. 2), the current in the vertical interface will get bent leftward and flows back as the outer-edge current on the upper edge. This current flow is also shown in Fig. 5(a). Fig. 5(c) is the transmission curve of this filter. We see that in the Fermi energy (E=0) region, the transmittance is 100%, which means all the K valley current is transmitted into the right part of the filter.

The lower panel of Fig. 5 is another valley filter: with QAH1|QAH2 in the left part and QVH1|QAH2 in the right part. The K-valley-polarized inner-current is blocked in the right inner-edge channel (see the band in Fig. 3, there is only one inner-edge band in K'-valley). So the K-valley component of the inner edge current is bent down in the vertical interface and then goes back on the lower edge of the left nanoribbon. Only K' valley current goes to the right lead with 100% transmittance (Fig. 5(f)).

Although there exist two outer-edge bands (states) on the left hybrid nanoribbon system (Fig. 2), we only observe one edge current in Fig. 5(b) or Fig. 5(e). The reason is that on the left/right interface region, the lower (upper) parts are of the same topological phase (QAH1 in Fig. 5(a), or QAH2 in Fig. 5(d)). So the vertical current only flows on the interface between different phases.

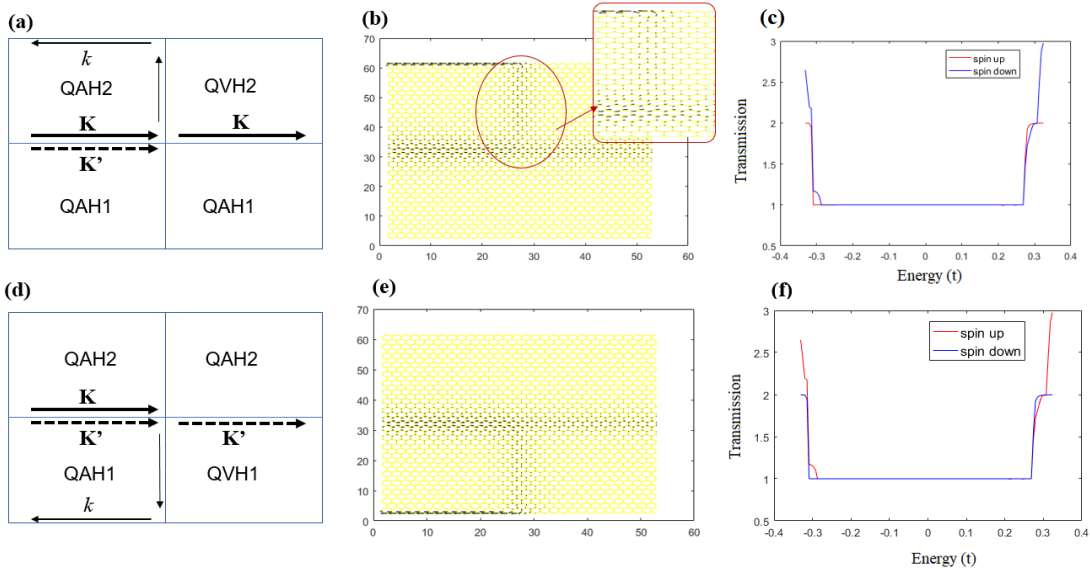

**Fig.5.** The topological phase and current demonstration ((a) and (d)), local current distribution ((b) and (e)) and the transmission spectra ((c) and (f)) of the inner-edge valley filter. The upper panel is of the topological structure QAH1|QAH2- QAH1|QVH2; the lower panel is of the topological structure QAH1|QAH2- QVH1|QAH2. In (a) and (d), the thick arrows denote the

valley-polarized current (solid line for K and dashed line for K') and the thin arrow denotes the non-valley-polarized current. In (b) the inset is the magnified current distribution near the interface of two hybrid parts.

**3.3 Valley-and-spin-polarized transport in hybrid nanoribbon junctions**

Now we design the filter for both valley and spin polarization based on these hybrid silicene-like nanoribbons. Fig. 6 shows the filter in the two-segments combination with the topological phases: QAH1|QAH2 in the left part and QAH1|SQVH1 in the right part.

Fig. 6(a) and (b) demonstrate the inner-edge current flows and the calculated local current distribution with spin-up case. Similar to Fig. 5(a), the K/K' valley polarized current exists in the left part and only K' valley polarized current can flow in the right part. So the K-valley current is blocked by the BWM effect as stated before. The blocked K-valley current goes upward along the interface of left/right nanoribbons and flows back as the non-valley polarized current along the upper edge. Here we also observe the valley-to-non-valley transformation along the vertical current region.

And Fig. 6(c) and (d) are the similar case but for the spin-down currents. We notice that besides the spin difference, the valleys of the filtered currents are also different: for spin up current it is in K' valley and for spin-down current it is in K valley. This is due to the SQVH1 topological phase. The band structure of QAH1|SQVH1 hybrid nanoribbon only has the spin up (down) edge band in K'(K) valley with the positive slope (Fig. 4). And in Fig. 6(a) and (c), we see that the lower part of the filter has the same phase (QAH1) and only the upper part has different phases across the left-right interface. This leads to the vertical current in the upper regions.

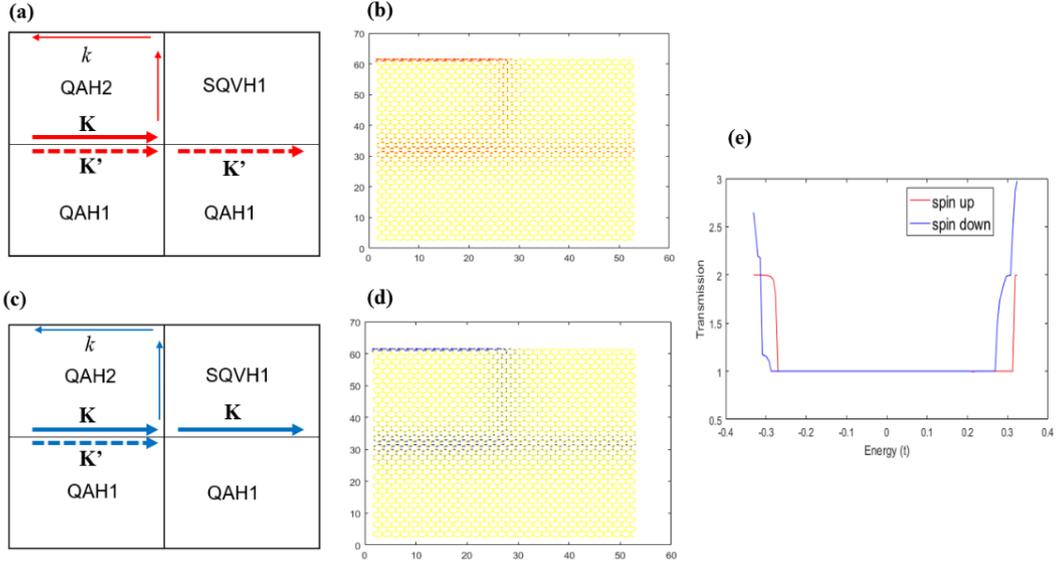

**Fig. 6.** The topological phase and current demonstration ((a) and (c)), local current distribution ((b) and (d)) and the transmission spectra ((e)) of the inner-edge valley filter (QAH1|QAH2-QAH1|SQVH1). In (a)-(d), the red arrows are for spin-up currents and blue arrows are for spin-down currents.

Figure 7 is the valley-spin filter with three segments. The left segment is of the QAH1|QAH2 type; the middle segment is of the QVH1|QAH2 type and the right segment is of the QAH1|SQVH1 type. As shown in the band structures of these segments, the left part has the inner-edge currents within K and K' valleys; the middle part supports the inner-edge state within the K valley; and the right part supports the inner-edge current within K' valley (spin up) or K valley (spin down). The local current distributions in Figs. 7(d) and 7(e) show that when the inner-edge currents go into the middle part, only K-valley-polarized current (spin up and spin down) continues to flow; the K'-valley-polarized currents are bent down to the lower outer edges of the ribbon. Then when the K-valley-polarized current goes into the right part, the spin-up current is bent and reflected along the upper edges; only the spin-down current continue to flow forwards with K-valley polarization. In other words, this filter has two layers of choosers: the first chooser lies in the interface between left and middle parts, which can select the valley-polarized incident inner-edge currents; the second chooser lies in the interface between the middle and right parts, which further selects the spin polarized

currents. Only the spin-down current within the K-valley can pass through this two-layer filter. Our calculated transmission is 97% for the spin-down current and 0.03% for the spin-up current. This filtering mechanism is also due to the BWM effect.

It is easy to see that with some changes in the external fields, we may obtain other filters with different valley-spin polarization.

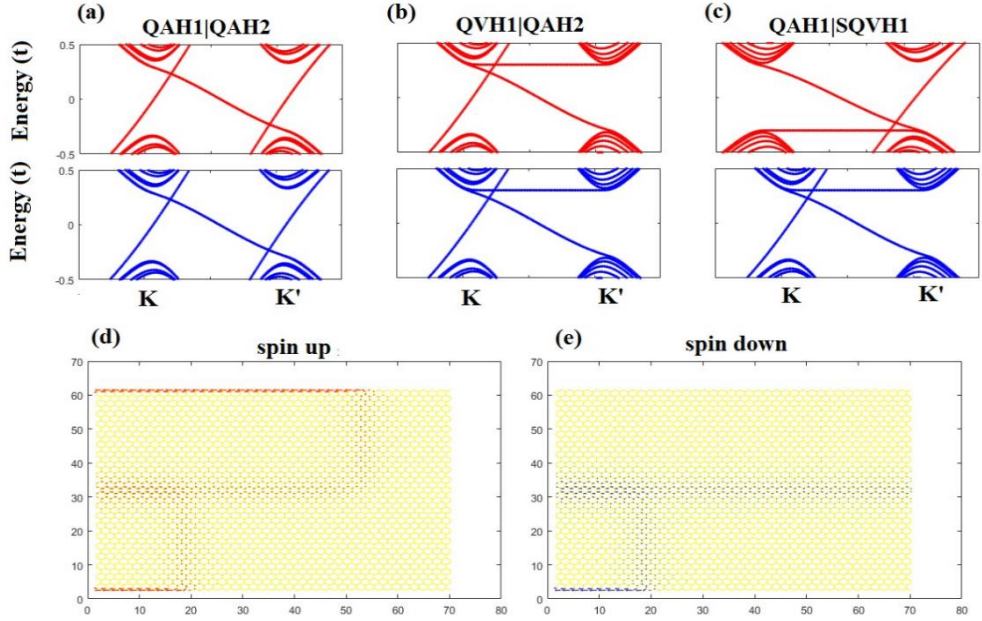

**Fig.7.** (a)-(c): The band structures of the three-segments filter (QAH1|QAH2- QVH1|QAH2- QAH1|SQVH1). The upper panel is for the spin-up bands and the lower panel is for the spin-down bands; (d) and (e): The local current distributions of this filter in the spin-up case ((d)) and the spin-down case ((e)).

## 4. Conclusions

With the tight-binding model, Jackiw-Rebbi interface theory and the NEGF theory, we investigate some inner-edge states with topological phases of QAH, QVH and SQVH based on the hybrid silicene-like nanoribbons. The band structures show some special inner-edge bands with valley and spin-valley polarization and different group velocities. And we find a new filtering mechanism called the Block wavevector mismatch. With this mechanism, we design the two-segments and three-segments serial inner-edge filters. These filters can generate valley-polarized or valley-spin-polarized inner-edge currents.

These inner-edge state can be fulfilled on experiments with different external fields exerted on the upper and lower parts of SiNR. For examples, people proposed that with different directions of vertical electric fields, the different AB staggered potentials appear on two sides of the inner-edge of SiNR and the corresponding inner-edge state form [10]; and with the radiation of the left-hand or right-hand circular polarized light on two regions of SiNR, the QAH1 and QAH2 states appear and the corresponding inner-edge state also form [35].

We believe these new inner-edge filters can be promising candidates in the future's valleytronic devices.

## Acknowledgements

We give thanks to the financial supports by the National Natural Science Foundation of China (Grant No. 12204073, 11847301), the Starting Foundation of Chongqing College of Electronic Engineering (Grant No.120727), the Guangxi Science and Technology Base and Talent Project (Grant No. 2022AC21077).

## Appendix Effect of the Rashba spin-orbital coupling

In the main text we do not consider the Rashba spin-orbital coupling. However, in some cases there exist this Rashba interaction in the materials. Here we give a brief investigation for this effect. The Rashba term reads $\sum_{<i,j>}\sum_{\alpha,\beta}\lambda_R c_{i\alpha}^{\dagger}(\boldsymbol{\sigma}\times\mathbf{d}_{i,j})|_{\alpha\beta}^{z} c_{i\beta}$. This term can mix spin up and down components of electron. We find that when the Rashba coupling strength ($\lambda_R$) is very small, the spin-polarized inner-edge state still remains. There is only a little opposite spin component which can be omitted; However, when the Rashba strength is large enough(for example, it is larger than $0.05t$), the two spin components are comparable and there is a mixed-spin state without any spin polarization. The following Fig. A1 gives this results for the inner-edge states in QAH1|SQVH1 case. And for the phase transitions of these topological phases, the similar case still exists: a small Rashba term does not change the phase transitions.

We note that if only the device region has this Rashba interaction, the eigen-state from the electrodes (without the Rashba term) is not the same as that in the device. So the spin-polarized current will be flipped into the opposite-spin one with increasing the transport length due to this Rashba effect. So in this case the Rashba term would destroy the spin-polarized current.

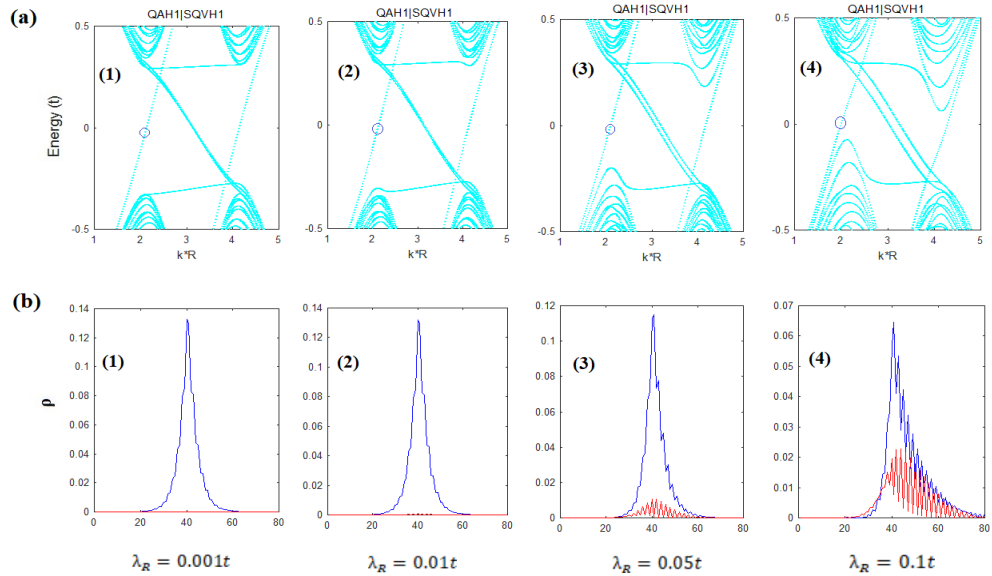

**Fig. A1**. Band structures (a) and electron distributions (b) of the inner-edge states (QAH1|SQVH1) with increasing Rashba intensities ($\lambda_R$). (1): $\lambda_R = 0.001t$; (1): $\lambda_R = 0.01t$; (3): $\lambda_R = 0.05t$; (4): $\lambda_R = 0.1t$. In (a) the blue circles indicate the band positions for the electron distribution calculations; In (b) the blue curves are for the spin down components and the red curves are for the spin up components of electron.